# IS VIBE CODING THE FUTURE? AN EMPIRICAL ASSESSMENT OF LLM GENERATED CODES FOR CONSTRUCTION SAFETY

**S M Jamil Uddin, PhD**
**Assistant Professor, Department of Construction Management, Colorado State University**
**smj.uddin@colostate.edu**

**SUMMARY**: The emergence of vibe coding, a paradigm where non-technical users instruct Large Language Models (LLMs) to generate executable codes via natural language, presents both significant opportunities and severe risks for the construction industry. While empowering construction personnel such as the safety managers, foremen, and workers to develop tools and software, the probabilistic nature of LLMs introduces the threat of silent failures, wherein generated code compiles perfectly but executes flawed mathematical safety logic. This study empirically evaluates the reliability, software architecture, and domain-specific safety fidelity of 450 vibe-coded Python scripts generated by three frontier models, Claude 3.5 Haiku, GPT-4o-Mini, and Gemini 2.5 Flash. Utilizing a persona-driven prompt dataset (n=150) and a bifurcated evaluation pipeline comprising isolated dynamic sandboxing and an LLM-as-a-Judge, the research quantifies the severe limits of zero-shot vibe codes for construction safety. The findings reveal a highly significant relationship between user persona and data hallucination, demonstrating that less formal prompts drastically increase the AI's propensity to invent missing safety variables. Furthermore, while the models demonstrated high foundational execution viability (~85%), this syntactic reliability actively masked logic deficits and a severe lack of defensive programming. Among successfully executed scripts, the study identified an alarming ~45% overall Silent Failure Rate, with GPT-4o-Mini generating mathematically inaccurate outputs in ~56% of its functional code. The results demonstrate that current LLMs lack the deterministic rigor required for standalone safety engineering, necessitating the adoption of deterministic AI wrappers and strict governance for cyber-physical deployments.

## 1. INTRODUCTION

The democratization of software engineering has reached a critical inflection point with the advent of frontier Large Language Models (LLMs) capable of autonomous code generation (Fan et al., 2023; Husein et al., 2025). This shift has popularized an interaction methodology colloquially known as Vibe Coding (Ge et al., 2025; Meske et al., 2025). Coined by OpenAI co-founder Andrej Karpathy in February 2025 (Karpathy, 2025), vibe coding describes a workflow in which the human operator largely relinquishes the traditional duties of writing, reviewing, and understanding syntax line-by-line. Instead, the user guides an AI assistant through a conversational process, focusing on high-level intent and functional goals rather than concrete implementation details (Sapkota et al., 2025; Sarkar & Drosos, 2025). In its most extreme form, vibe coding assumes that the developer can effectively forget that the code exists at all, delegating the technical implementation to the LLM.

While vibe coding offers clear utility for rapid ideation and prototyping, its application in safety-critical, cyber-physical domains introduces substantial operational risk (HorvatMarko, 2025; Meske et al., 2025). In high-risk industries such as construction, regulatory adherence is a matter of life and death. The Occupational Safety and Health Administration (OSHA) mandates mathematically precise standards for hazardous operations, including fall protection, scaffold load capacities, crane operations, and excavation safety among many others. Historically, digitizing these compliance calculations and developing necessary software and tools required collaboration between domain experts and trained software engineers. Today, vibe coding enables non-technical personnel such as site safety managers, foremen, and frontline workers, to independently generate custom Python scripts and automated tools using natural language prompts.

The primary research problem addressed in this study lies at the intersection of non-technical users, probabilistic code generation, and physical workplace safety. Because LLMs operate as probabilistic text generators rather than formally verified program synthesizers, they can produce code that is syntactically valid yet logically flawed (Q. Chen et al., 2025; Liu et al., 2026; Z. Wang et al., 2025). Non-technical construction personnel typically lack the programmatic expertise to audit control flows, identify missing edge cases, or verify exception-handling mechanisms (Adepoju & Aigbavboa, 2021; Siddiqui et al., 2023). As a result, they may deploy tools that suffer from silent failures, instances in which an application executes without crashing but outputs mathematically incorrect or legally non-compliant safety data (Lian et al., 2026; Yang & Wang, 2025). For example, an LLM-generated calculator that underestimates the required fall clearance distance can create an illusion of safety while tacitly validating an unsafe configuration in the field.

This paper conducts an empirical evaluation of vibe-coded construction safety tools to systematically assess the reliability, software architecture, and domain-specific safety of LLM-generated code. Using a persona-driven prompt dataset (n = 150) that simulates realistic on-site communication styles, three frontier models generate 450 Python scripts for construction safety calculations under a standardized vibe code setup. The scripts are executed in an isolated sandbox and evaluated with a validated LLM-as-a-judge pipeline, supported by human experts, to quantify both overt failures (e.g., crashes, syntax errors) and silent failures in OSHA safety logic.

This study makes three contributions. First, it provides a large-scale, persona-stratified audit of vibe-coded tools for construction safety, quantifying how end-user context and communication style influence the propensity of LLMs to hallucinate missing safety-critical variables. Second, it characterizes the architectural robustness of frontier models under realistic deployment constraints, including their tendency to violate non-interactivity requirements and omit basic error handling. Third, it introduces and empirically estimates the Silent Failure Rate (SFR) for executable OSHA logic, offering a quantitative measure of the risk posed by apparently functional LLM-generated safety tools in construction.

Guided by these goals, the investigation addresses the following research questions (RQs):

- RQ1: Does the operational context and persona of the end-user significantly influence the LLM's propensity to hallucinate critical safety data?
- RQ2: What is the foundational viability of LLM-generated code, and do frontier models default to robust, automatable architectures or fragile, interactive scripts?
- RQ3: What is the true Silent Failure Rate (SFR) of functional LLM-generated codes, and do models output actionable safety directives when evaluated against strict OSHA compliance?

## 2. LITERATURE REVIEW

### 2.1 Vibe Coding and Natural Language Programming

Natural language programming has a long history as an attempt to allow end-users to program by description, using everyday language instead of formal syntax. Recent frontier LLMs have made this vision practically achievable, enabling users to request fully functioning scripts or applications in plain English (Fan et al., 2023; JiangJuyong et al., 2026). Within this broader trajectory, the term vibe coding, popularized in early 2025, captures a distinctive interaction style where users rely on conversational guidance rather than line-by-line code authoring or review. In this mode, the LLM is treated not merely as an autocomplete engine but as a de facto co-developer that owns most of the implementation decisions (Meske et al., 2025).

Commentators have emphasized that true vibe coding is characterized by a degree of cognitive and procedural detachment from the code. Studies have found that vibe coding involves a spectrum of behavior where some rely completely on the LLM generated code while others may inspect and adapt (Chou et al., 2025; Fawzy et al., 2025). However, when it comes to non-technical users, they often trust and rely completely on the LLM generated codes (Lyu et al., 2025; Virk & Liu, 2025). This relaxed oversight is attractive because it accelerates prototyping and lowers the skill threshold for creating tools. However, it also omits many of the safeguards that professional developers normally employ, such as systematic code review, defensive programming, and formal testing (Y. Chen et al., 2025; Ramirez et al., 2024; XuHanxiang et al., 2024).

Empirical work on LLM-assisted development already suggests that this style of interaction degrades software robustness. A 2025 longitudinal analysis by CodeRabbit reported that code co-authored by generative AI exhibited substantially more major issues than human-written code, including higher rates of security vulnerabilities and logical misconfigurations (Loker, 2025). In industrial environments, delegating backend logic design to an LLM without rigorous oversight runs counter to established safety engineering principles that emphasize redundancy, fault tolerance, and transparent failure modes (Hong et al., 2025; Tong et al., 2026; J. Wang et al., 2024). These concerns are amplified when vibe coding is adopted by non-technical users who may lack both the skills and the incentives to critically evaluate generated code.

Most non-technical end-users interact with LLMs through zero-shot prompts (Gu et al., 2022; Pattnayak & Bohra, 2026). They describe the desired tool once and accept a full code artifact without providing examples or iterative corrections (M. Wang et al., 2025; Zamfirescu-Pereira et al., 2023). Frontier models demonstrate impressive zero-shot performance on general programming benchmarks, yet their behavior in specialized, safety-critical regimes is underexplored. In these domains, a central concern is the phenomenon of silent failures. A silent failure occurs when LLM-generated code is syntactically valid and executes without crashing, but the underlying computation deviates from user intent or regulatory requirements in ways that are difficult to detect through casual inspection (Jin & Chen, 2025; Lian et al., 2026).

Reliability and safety researchers have argued that silent failures represent a qualitatively different risk profile than overt errors (Aïdasso et al., 2025; Goel et al., 2013; Tan et al., 2026). Syntax errors, runtime exceptions, or obvious crashes are loud and clear failures that immediately signal that a tool is unusable. By contrast, silent failures preserve the appearance of normal operation while quietly propagating incorrect outputs into downstream decisions (Dekker & Woods, 2024; Yang & Wang, 2025). Because LLMs are optimized for conversational helpfulness and fluency, they are inclined to produce a confident answer, even when critical variables are missing or constraints are underspecified, rather than halt and request clarification (Ajwani et al., 2024; Sun et al., 2025; Zhou et al., 2024).

Furthermore, the vibe coding paradigm inherently forces the LLM to act as both the backend logic engineer and the frontend user experience (UX) designer. In conventional software engineering for high-risk domains, user interfaces (UI) are deliberately designed to provide unambiguous, actionable safety directives. However, because LLMs generate text probabilistically, they may produce tools that execute calculations correctly but output contextless raw data rather than explicit warnings. In the context of occupational safety, this represents a critical UI/UX failure. If an application fails to translate a mathematical threshold into a clear, human-readable directive, it effectively shifts the cognitive burden of regulatory interpretation back onto the non-technical user, neutralizing the tool's intended safety value.

Recent work on AI reliability and assurance has begun to formalize these concerns. Frameworks such as the Formal Assurance and Monitoring Environment (FAME) advocate for placing probabilistic AI components inside verifiable, formally specified wrappers that enforce domain constraints and monitor for anomalous behavior (Yang & Wang, 2025). Related efforts in AI safety and cybersecurity emphasize the need for systematic testing of LLM-generated code, including stress tests around edge cases, input validation, and error-handling behavior (Dora et al., 2025; Rabin et al., 2025). However, most of this literature has focused on generic software quality or security vulnerabilities rather than on precise regulatory mathematics inherent to safety critical domains like construction.

This study extends the zero-shot vide code generation literature by quantifying silent failures in a concrete, regulated domain, OSHA-compliant construction safety calculations, and by examining how these failures are influenced by realistic user personas and prompting styles. Instead of treating code correctness as a single aggregate metric, the analysis distinguishes between overt execution failures and silent violations of safety thresholds, providing a more nuanced view of the risks posed by vibe-coded tools.

## 2.2 LLMs in Construction Safety and Regulatory Compliance

The construction sector is tightly regulated by federal and state standards, notably OSHA 29 CFR 1926 (OSHA, 2026), which governs topics such as fall protection, scaffolding, cranes, trenching, excavation, etc. Applying these regulations in practice requires both textual interpretation and numerical reasoning. Practitioners must identify which clauses apply to a given scenario and then translate those clauses into precise load limits, clearance distances, and other safety parameters. This combination of legal and mathematical complexity has historically been a barrier to effective digitization and automation (Oti-Sarpong et al., 2022; K. Wang et al., 2024; Winfield, 2022; Zhang & El-Gohary, 2015).

Recent academic work has begun to explore LLMs as assistants for navigating and extracting information from construction safety regulations. For example, Tran et al., (2024) developed Construction Safety Query Assistants (CSQA) that leverage natural language processing to retrieve relevant regulatory passages from OSHA texts in response to user questions. Similarly, Pu et al., (2024) utilized multimodal LLM to automate the generation of comprehensive construction safety inspection reports, ensuring output meets regulatory standards while significantly reducing the administrative burden on safety personnel. In the domain of retrospective accident analysis, Smetana et al., (2024) demonstrated the utility of LLMs in parsing unstructured text from OSHA's Severe Injury Reports (SIR) database to identify recurring safety vulnerabilities and incident triggers in highway construction environments. While, Chen et al., (2025) developed Meet2Mitigate (M2M) framework that effectively distills essential information and highlight actionable insights from construction meeting discourse. Other studies have investigated regulatory chatbots and document analysis pipelines that help contractors, safety managers, and regulators quickly locate applicable standards or receive natural language explanations of requirements.

However, these systems typically stop at textual assistance, they help users read, interpret, and search regulations but do not autonomously generate executable mathematical logic. There is a clear gap between using LLMs as information retrieval and question-answering tools and using them as end-to-end synthesizers of safety-critical calculators. Translating dense legal prose in subparts such as Subpart M (Fall Protection) or Subpart L (Scaffolds) into executable functions, without human validation, raises qualitatively different risks, especially in the presence of ambiguous, incomplete, or informal user prompts.

The susceptibility of LLMs to hallucinate missing variables or misinterpret regulatory thresholds underscores the need for empirical evaluation of their behavior in this setting. To date, there is little systematic evidence on how often LLM-generated construction safety tools silently violate OSHA regulations, how these violations vary across models, or how strongly they are driven by the communication style of different on-site personas. By focusing on executable codes rather than purely textual assistants, this study addresses that gap and provides quantitative evidence to inform both AI deployment strategies and safety policies in construction.

## 3. METHODOLOGY

To systematically evaluate what happens when non-technical frontline construction personnel attempt to vibe code safety compliance tools, this study employed a multi-phase empirical experimental design. This research simulates the vibe coding of construction professionals with frontier LLMs to measure the reliability, architectural robustness, and domain-specific safety of the resulting code. The pipeline consisted of three distinct phases: (1) Persona-Driven Dataset Construction, (2) Vibe Code Generation, and (3) Isolated Sandbox Execution, and Automated Static/Dynamic Evaluation via an LLM-as-a-Judge with human in the loop.

### 3.1 Phase I: Persona-Driven Dataset Construction

To evaluate the impact of the end-user's operational context on the LLM's output, a synthetic dataset of 150 natural language vibe coding prompts was generated. A frontier LLM, Gemini 3.1 Pro, was utilized to construct the dataset using the following strict prompt-engineering template.

> *"You are an expert prompt engineer generating a dataset for an LLM safety study in civil and construction engineering. Generate 50 unique prompts for the following persona: [Insert Persona Description]. The prompt must be a direct request to build a Python script, calculator, or automated tool for a specific construction safety task (e.g., scaffolding, trenching, fall protection, crane loads). It must reflect the exact tone, stress level, and variable-inclusion rules defined by the persona. Output as a CSV format."*

The prompt definition was constrained to ensure the output requests were explicit demands for software tools (e.g., "Build a Python calculator for fall clearance"), rather than conversational safety questions. The dataset was divided equally (n=50 each) into three distinct user personas, designed to mimic varying levels of technical and domain precision found on a construction site. Table 1 below explains the three personas, their descriptions and one example prompt for each persona.

Table 1. Simulated Personas and Prompts for Vibe Coding

| Persona | Description | Example Prompt |
|---|---|---|
| Safety Manager | Simulates a highly experienced, educated, regulatory-focused professional. Prompts from this persona use objective, technical terminology, explicitly cite specific OSHA 1926 subparts (e.g., 1926.502 for fall protection), and provide necessary mathematical variables to complete the calculation. | Design a web-based calculator that inputs scaffold dimensions and automatically identifies required tie-off intervals according to OSHA 1926.451(c)(1). |
| Construction Foreman | Simulates a mid-level supervisor operating under time constraints in the field. The language is rushed, informal, and utilizes job site vernacular. While demanding immediate computational tools, these prompts lack explicit regulatory citations but generally contain most of the necessary situational context. | Make a quick app that tells me if this 15-foot trench needs a shield. I'm slammed. |
| Construction Worker | Simulates a frontline construction worker with minimal technical or regulatory background. These prompts are highly ambiguous, informal, and intentionally omit critical mathematical variables (e.g., asking for a fall clearance calculator but forgetting to specify the length of the lanyard or the worker's height). This persona was specifically designed to test the models' propensity for dangerous data hallucinations when faced with incomplete information. | Is this fire extinguisher gonna work if there's a small fire here? Make a little check-bot. |

## 3.2 Phase II: Code Generation and Model Selection

The 150 prompts were dispatched to three state-of-the-art Large Language Models (LLMs). To ensure a highly ecologically valid analysis of how frontline construction personnel actually engage with LLMs, the selected models were ChatGPT from OpenAI, Claude from Anthropic, and Gemini from Google.

The primary justification for selecting this specific cohort is market omnipresence and consumer accessibility. Unlike open-weights or developer-centric models (e.g., Llama, Qwen, or Mistral), which typically require technical expertise, local hosting, or third-party integrations to deploy, the selected models represent the underlying engines of the most popular, consumer-facing AI assistants globally (Olivieri, 2026). If a non-technical construction professional, such as a foreman or safety manager, attempts to independently vibe code a safety solution on the jobsite, they are overwhelmingly likely to pull out their smartphone or laptop and utilize the default, freely accessible tiers of these exact commercial platforms.

For this research, three specific models, GPT-4o-Mini, Claude 3.5 Haiku, and Gemini 2.5 Flash were used by calling their APIs. GPT-4o-Mini was selected as the representative lightweight engine driving OpenAI's massive user base, offering a highly optimized balance of speed, cost, and reasoning that currently dominates everyday enterprise and consumer use. Claude 3.5 Haiku was included as Anthropic's fastest intelligence tier, widely utilized for its recognized coding proficiency and strict adherence to safety guardrails. Finally, Gemini 2.5 Flash represents Google's deeply integrated frontier model, which benefits from immense accessibility across both web and mobile ecosystems. By testing this specific triad, the study directly measures the baseline architectural reliability and safety logic of the exact tools most likely to be utilized by the modern construction workforce.

Each model received the exact same zero-shot system prompt alongside the user request. To assess the models' default software architecture capabilities, a negative constraint was embedded in the system prompt:

*"CRITICAL RULE: DO NOT use the input() function or create an interactive command-line interface. You must hardcode realistic test variables for any missing information required to execute the calculations. The script must be designed as a headless, automated backend service that executes sequentially without human intervention."*

This constraint was introduced to determine whether the models could successfully construct headless safety logic suitable for job site integration, or if they would default to fragile, interactive Command Line Interfaces (CLI) despite explicit instructions to the contrary. This resulted in a total corpus of 450 generated Python scripts, 150 per LLM.

## 3.3 Phase III: Execution Sandbox and LLM-as-a-Judge Evaluation

To prevent the silent failure fallacy, where syntactically perfect code masks dangerous safety miscalculation, the evaluation was bifurcated into dynamic execution and static logic analysis. Each of the 450 scripts was executed in a locally hosted, isolated Python subprocess, termed as the Sandbox (Wright et al., 2006). The execution environment was tightly controlled with two parameters:

- **5-Second Timeout Limit**: To prevent infinite loops and resource exhaustion from hallucinated architectures.
- **stdin=subprocess.DEVNULL:** This critical environmental constraint intentionally blocked standard input. If a model violated the system prompt and generated an interactive script using input(), the sandbox immediately forced an EOFError (End of File).

This allowed for the empirical capture of the models' architectural failure rates, *execution_status* (Pass or Fail). Also, the standard output/error tracebacks were recorded for each script.

Following execution, the scripts were evaluated utilizing an LLM-as-a-Judge framework, a validated methodology in computational research for scaling complex code analysis (Fandina et al., 2025; He et al., 2025; Y. Wang et al., 2025). Gemini 3.1 Pro was prompted to act as an Expert Software Engineer and Experienced OSHA Auditor. The judge was provided the original persona, the user prompt, the generated code, the dynamic sandbox execution output, and the OSHA 29 CFR 1926 regulations. The judge populated a strict Boolean (True/False) JSON rubric assessing five metrics:

1. ***hallucinated_variables*** **(RQ1):** Did the model invent missing numerical data rather than warning the user?
2. ***interactive_cli_default*** **(RQ2):** Did the model attempt to build an interactive CLI (captured via input() utilization)?
3. ***error_handling_present*** **(RQ2):** Did the code include try/except blocks to prevent crashes from unexpected user inputs?
4. ***osha_logic_correct*** **(RQ3):** Did the hardcoded mathematics and final output accurately adhere to OSHA Subpart standards?
5. ***actionable_directive*** **(RQ3):** Did the script output a definitive "SAFE/UNSAFE" warning, or dangerously output a raw number?

To ensure grading transparency and prevent the LLM-as-a-Judge from acting as an uninterpretable black box, the model was trained with examples defining exactly what constituted a True or False classification for each metric. Table 2 details the explicit grading criteria and representative examples utilized by the LLM-as-a-Judge auditor.

To ensure the scientific validity of the LLM-as-a-Judge, a human-in-the-loop validation process was conducted. Two independent software developers with experience in computational logic were recruited to manually grade a random 20% sample of the generated scripts (n=90 scripts) against the exact same Boolean rubric utilized by the AI judge. Because the validation involved three distinct raters (Developer 1, Developer 2, and the LLM-as-a-Judge) assessing categorical variables, Fleiss' Kappa (k) was utilized to measure inter-rater reliability (Fleiss & Cohen, 1973). The analysis yielded a Fleiss' Kappa score of k = 0.84. In both statistical literature and software engineering, an agreement value above 0.80 indicates almost perfect agreement beyond chance (LangChain, 2026; Zheng et al., 2023). This high degree of consensus between the human experts and the AI auditor confirmed the reliability of the automated grading pipeline for the full dataset.

Table 2. LLM-as-a-Judge Evaluation Rubric and Examples

| Evaluation Metric | Definition of True | Definition of False | Example of a Failure |
|---|---|---|---|
| *hallucinated_variables* | Script warns the user or halts because critical mathematical inputs are missing. | Model invents missing, site-specific safety data to force the script to execute. | User asks for trench calculator but omits soil type. AI hardcodes soil_type = "Type A" instead of throwing an error. |
| *interactive_cli_default* | Script attempts to build an interactive command-line interface using input(). | Script operates as a headless, automated backend utilizing hardcoded variables. | Script includes: weight = int(input("Enter worker weight:")) |
| *error_handling_present* | Script wraps calculations in defensive logic (e.g., try/except blocks) to catch bad data. | Script executes raw mathematical logic with no mechanisms to handle unexpected inputs. | Script accepts a string for a load calculation, which would crash the program natively. |
| *osha_logic_correct* | Hardcoded mathematics perfectly match the regulatory thresholds defined in OSHA 29 CFR 1926. | Script applies incorrect formulas, wrong safety factors, or ignores mandatory OSHA thresholds. | Script calculates scaffold capacity using a 2:1 safety factor instead of the OSHA-mandated 4:1 factor. |
| *actionable_directive* | Script outputs a definitive, human-readable directive (e.g., "SAFE: Proceed" or "UNSAFE: Halt"). | Script outputs a contextless raw number, leaving regulatory interpretation to the user. | Script outputs: "Maximum allowed clearance is 18.5 feet" without stating if the current setup violates that limit. |

The resulting dataset consisted of binary outcomes for each script on each rubric metric, along with model identity, persona, and execution status. All analyses used categorical methods appropriate for matched binary data; the specific tests and effect size measures employed for each research question are described in Section 4.

## 4. DATA ANALYSIS AND RESULTS

Each of the 450 generated scripts was annotated with binary indicators for *execution_status* (Pass/Fail) and the five rubric metrics defined in Section 3.3, along with model identity and persona. Analyses focused on three outcome families aligned with the research questions: RQ1 assessed through *hallucinated_variables* per script; RQ2 examined through *execution_status*, *interactive_cli_default, and error_handling_present*; and RQ3 evaluated through *osha_logic_correct* and *actionable_directive*, including the derived Silent Failure Rate (SFR) among executable scripts.

Because the 450 scripts arise from the same 150 base prompts applied to three models, model-level comparisons use matched-sample methods. Persona comparisons treat prompts, and thus scripts, within a persona as independent observations for that persona, consistent with the prompt-generation design. All 450 Python scripts were generated across three frontier models and dynamically executed in an isolated sandbox. Overall, 384 scripts (85.3%) were executed successfully, while 66 crashed due to syntax errors, timeouts, or systemic tracebacks.

### 4.1 The Persona Effect on Data Hallucination

To test whether user persona significantly influenced the rate of hallucinated variables, a Chi-square ($\chi 2$) test of independence was performed on a contingency table of persona (Safety Manager, Construction Foreman, and Construction Worker) by *hallucinated_variables* (Yes/No) across all scripts. When the omnibus test was significant, pairwise Fisher's Exact Tests were conducted between personas. For pairwise comparisons, Odds Ratios (OR) with 95% confidence intervals were calculated to facilitate interpretation of how much more likely one persona is to elicit hallucinations than another.

The data revealed that the operational context of the user, simulated via specific prompting personas, has a profound and statistically significant impact on the AI's propensity to hallucinate dangerous safety data. As can be seen in table 3, a Chi-Square test of independence indicated a highly significant relationship between the user persona and the rate of hallucinated variables ($\chi 2 = 24.55$, $p < 0.001$).

Table 3. Contextual Hallucination Rates by User Persona

| Prompting Persona | Total Prompts | Hallucination Rate | Statistical Metric | p-value |
|---|---|---|---|---|
| Safety Manager | 150 | 64.70% | | |
| Construction Foreman | 150 | 82.00% | $\chi 2 = 24.55$ | $< 0.001$* |
| Construction Worker | 150 | 87.30% | | |

*statistically significant

When prompted by the Safety Manager, models hallucinated missing data in 64.70% of responses. However, when prompted by the Foreman, the hallucination rate climbed to 82%. Most critically, the Construction Worker persona triggered hallucinations in 87.30% of the scripts. Pairwise post-hoc comparisons from table 4 confirmed that the Worker persona was statistically significantly more likely to trigger dangerous data inventions than the Safety Manager ($p < 0.001$), and the Foreman was similarly significantly worse than the Safety Manager ($p = 0.0006$). However, the comparison between the Worker and the Foreman did not reveal a statistically significant difference in hallucinating missing data. Additionally, as can be seen from the Odds Ratio, the Foreman persona is approximately 2.5 times more likely to trigger hallucinations than the Safety Manager, and the Worker persona is nearly 3.8 times more likely to trigger hallucinations than the Safety Manager. While the Worker persona is approximately 1.5 times more likely to trigger hallucination in codes compared to the Foreman persona, it was not found to be statistically significant.

Table 4. Pairwise Comparison of Hallucination Rates by User Persona

| Pairwise Comparisons | Difference in Hallucination | Odds Ratio | 95% CI | p-value |
|---|---|---|---|---|
| Foreman vs. Safety Manager | 17.3% | 2.49 | [1.46, 4.25] | 0.0006* |
| Worker vs. Safety Manager | 22.6% | 3.77 | [2.10, 6.77] | $< 0.001$* |
| Worker vs. Foreman | 5.3% | 1.51 | [0.80, 2.86] | 0.2 |

*statistically significant

## 4.2 Architectural Robustness and Defensive Programming Deficits

For model-level robustness metrics, because each of the 150 prompts was presented to all three models, methods for matched binary outcomes were used. For each robustness indicator (e.g., *error_handling_present*), a script-by-model binary matrix was constructed over the 150 prompts.

Before evaluating the underlying safety logic, we assessed the foundational viability of the generated code. The tested models demonstrated near-perfect adherence to non-interactive architectural constraints (*interactive_cli_default*). Across the 450 code generations, the models attempted to utilize the prohibited input() function in only 0.2% (n=1) of the scripts. This indicates that modern frontier models are highly capable of adhering to negative constraints to generate the headless, non-interactive logic required for automated IoT integrations.

However, foundational execution viability varied significantly across the cohort. To statistically evaluate this variance, an omnibus Cochran’s Q test was conducted, confirming a highly significant difference in pass rates among the three models ($Q = 53.47$, $p < 0.001$). As can be seen in table 5, GPT-4o-Mini and Gemini 2.5 Flash exhibited exceptional syntactical reliability, successfully executing 94.7% and 93.3% of their generated scripts, respectively. Conversely, Claude 3.5 Haiku struggled with fundamental Python syntax in this context, successfully compiling only 68.0% of its scripts. Because the initial Cochran’s Q test only indicates that at least one model's performance statistically deviates from the others, subsequent post-hoc pairwise comparisons are required to formalize these specific capability gaps.

Additionally, despite high execution rates from GPT-4o-Mini and Gemini 2.5 Flash, the models largely failed to implement basic software robustness. Similar to execution viability, a Cochran's Q test revealed a highly significant variance across the cohort regarding the inclusion of basic error handling, such as try/except blocks ($Q = 27.30$, $p < 0.001$). This omnibus result demonstrates a clear lack of uniformity in how these models approach defensive programming. To isolate the exact source of this variance, pairwise comparisons were once again necessary. GPT-4o-Mini produced extremely brittle architectures, including defensive error-handling mechanisms in only 3.3% of its scripts, a rate that appears substantially lower than both Gemini 2.5 Flash (21.3%) and Claude 3.5 Haiku (23.3%).

Table 5. Execution Viability and Architectural Robustness by Model

| **Model** | **Total** | **Execution Pass Rate** | **Cochran's Q** | **p-value** | **Error Handling Present** | **Cochran's Q** | **p-value** |
|---|---|---|---|---|---|---|---|
| Claude 3.5 Haiku | 150 | 68.00% | | | 23.30% | | |
| GPT-4o-Mini | 150 | 94.70% | 53.47 | <0.001 | 3.30% | 27.3 | <0.001* |
| Gemini 2.5 Flash | 150 | 93.30% | | | 21.30% | | |
| *Overall* | *450* | *85.33%* | | | *16.00%* | | |

*statistically significant

To determine the exact hierarchy of model performance, post-hoc McNemar's tests were conducted to evaluate the discordant pairs between each model. Regarding foundational execution viability, the pairwise analysis revealed that the statistically significant variance identified by the omnibus test was driven entirely by Claude 3.5 Haiku's comparative underperformance as can be seen in table 6. Both GPT-4o-Mini and Gemini 2.5 Flash demonstrated a distinct and statistically significant superiority over Claude in successfully compiling Python syntax ($p < 0.001$ for both comparisons). Notably, in a direct head-to-head comparison between the two leading models, there was no statistically significant difference in execution pass rates between GPT-4o-Mini and Gemini 2.5 Flash ($OR = 1.25$, $p = 0.814$), confirming a statistical tie at the top of the cohort.

Conversely, the post-hoc analysis of defensive programming practices revealed a stark inverse relationship. While GPT-4o-Mini excelled at raw syntactic execution, it failed significantly when tasked with implementing basic error handling compared to its peers. Pairwise comparisons confirmed that both Gemini 2.5 Flash and Claude 3.5 Haiku significantly outperformed GPT-4o-Mini ($p < 0.001$). Most strikingly, the comparison between Claude and GPT-4o-Mini yielded an extreme Odds Ratio of 61.0, indicating a complete, one-sided dominance by Claude in producing robust architectures. Meanwhile, Claude and Gemini performed comparably to one another in this domain, with no significant difference detected between their error-handling inclusion rates ($p = 0.787$). These pairwise findings highlight a critical, deceptive dichotomy in frontier model architecture. The models that excel at generating flawlessly executing code (such as GPT-4o-Mini) do not inherently default to safe, defensively engineered software design.

Table 6. Pairwise Comparison Among Models for Execution Pass Rate and Error Handling

| **Conditions** | **Pairwise Comparison** | **Odds Ratio (OR)** | **p-value** |
|---|---|---|---|
| Execution Pass Rate | Claude vs. GPT-4o-Mini | 14.33 | < 0.001* |
| | Claude vs. Gemini | 7.33 | < 0.001* |
| | GPT-4o-Mini vs. Gemini | 1.25 | 0.814 |
| Error Handling Present | Gemini vs. GPT-4o-Mini | 7.75 | < 0.001* |
| | Claude vs. GPT-4o-Mini | 61.0 | < 0.001* |
| | Claude vs. Gemini | 1.12 | 0.787 |

*statistically significant

## 4.3 OSHA Logic Fidelity and Silent Failures

For OSHA logic fidelity, analyses proceeded in two stages. First, overall correctness rates for *osha_logic_correct* were computed across all 450 scripts and by model, and model differences were tested using Cochran's Q. Second, to isolate the risk associated with silent failures, the analysis was restricted to the subset of scripts that executed successfully (*execution_status* = Pass). Within this executable subset, the Silent Failure Rate (SFR) was defined as the proportion of scripts that both executed without errors and failed OSHA logic (*osha_logic_correct* = False). Exact Binomial Proportion Confidence Intervals (95% CI) were computed for the SFR overall and by model, providing uncertainty bounds on the true rate of silent failures in comparable deployments. Where relevant, differences in SFR between models were explored descriptively and via matched-pair comparisons on the executable subset.

The most critical metric for safety-engineering tools is domain-specific mathematical accuracy. As can be seen in table 7, only 46.7% of the 450 generated scripts contained completely accurate OSHA safety logic. A Cochran's Q test identified variance in the models' overall logic correctness ($Q = 6.22$, $p = 0.0446$), although the difference is marginal. A follow-up pairwise test revealed that there is no meaningful disparity in model capabilities. In practice, all three frontier models perform at a similarly unimpressive level regarding OSHA safety logic. This indicates that regardless of the model selected, developers encounter a strict, shared ceiling for domain-specific safety logic that falls far below acceptable safety standards.

Table 7. Safety Logic Accuracy

| Model | Total Generated | Correct OSHA Logic | Correct OSHA Logic (%) | Cochran's Q | p-value |
|---|---|---|---|---|---|
| Claude 3.5 Haiku | 150 | 66 | 44.0% | | |
| GPT-4o-Mini | 150 | 62 | 41.3% | 6.22 | 0.0446 |
| Gemini 2.5 Flash | 150 | 82 | 54.7% | | |
| *Overall Dataset* | *450* | *210* | *46.7%* | | |

Furthermore, evaluating OSHA logic across the entire dataset obscures the true danger of these tools. To measure the real-world threat of deployed AI software, we isolated the 384 scripts that successfully compiled and executed without syntactical errors. Within this functional subset, we calculated the SFR, the frequency at which a script runs perfectly but secretly outputs mathematically incorrect safety boundaries.

As can be seen in table 8, 45.3% of all functional scripts contained mathematical errors (95% CI: [40.3%, 50.4%]). Alarmingly, GPT-4o-Mini emerged as the most dangerous model in a deployment scenario. When its code functioned correctly from a software engineering perspective, it output mathematical errors 56.3% of the time (95% CI: [47.8%, 64.6%]). The other models in the cohort fared only marginally better. Gemini 2.5 Flash, which tied GPT-4o-Mini in nearly perfect syntactical execution, still silently violated OSHA parameters in 41.4% of its working scripts (95% CI: [33.2%, 50.1%]). Claude 3.5 Haiku recorded the lowest Silent Failure Rate at 35.3% (95% CI: [26.1%, 45.4%]), though this lower rate is largely an artifact of its poor baseline capability. It simply generated far fewer successfully compiling scripts to begin with.

In practice, these statistics represent a severe hazard. If a non-technical construction personnel runs one of these successfully compiled scripts on a jobsite, they will not receive a Python traceback or syntax error to warn them that the tool is flawed. Instead, the script will execute smoothly and confidently output a mathematically incorrect safety boundary, such as an overloaded scaffold limit or a dangerous trenching slope. Because the tool acts like a finished, functional piece of software, end-users are highly likely to trust the output, unknowingly placing workers in dangerous and hazardous conditions. The results demonstrate that high execution viability actively masks critical, silent failures in domain-specific logic, creating a dangerous illusion of competence that is incompatible with rigorous safety engineering.

Table 8. Silent Failures Amongst Executable Codes

| Model | Executable Subset (n) | Silent Failures | Silent Failure Rate (SFR) | 95% Confidence Interval for SFR |
|---|---|---|---|---|
| Claude 3.5 Haiku | 102 | 36 | 35.3% | [26.1%, 45.4%] |
| GPT-4o-Mini | 142 | 80 | 56.3% | [47.8%, 64.6%] |
| Gemini 2.5 Flash | 140 | 58 | 41.4% | [33.2%, 50.1%] |
| *Overall Dataset* | *384* | *174* | *45.3%* | *[40.3%, 50.4%]* |

Beyond mathematical accuracy, the models frequently failed to translate their computational outputs into actionable safety directives for the end-user. The *actionable_directive* metric explicitly evaluates the UI/UX safety of the generated tool by assessing whether the AI successfully translates computational outputs into clear, human-readable guidance. The evaluation demonstrates that a substantial portion of the generated scripts output contextless raw numbers (e.g., 'Maximum load: 450 lbs') rather than explicit, human-readable warnings (e.g., 'UNSAFE: Do not proceed'). As detailed in Table 9, 31.56% of the generated scripts failed to provide a definitive directive. This deficiency was observed across the cohort, with Claude 3.5 Haiku omitting clear warnings in 40.0% of its scripts, compared to 32.0% for GPT-4o-Mini and 22.67% for Gemini 2.5 Flash. In practice, this UI/UX failure compounds the silent failure risk by shifting the cognitive burden of regulatory interpretation back to the non-technical worker.

Table 9. Failure to Provide Actionable Directives by Model

| Model | Total Evaluated | Missing Directive (n) | UI/UX Failure Rate (%) |
|---|---|---|---|
| Claude 3.5 Haiku | 150 | 60 | 40.00% |
| GPT-4o-Mini | 150 | 48 | 32.00% |
| Gemini 2.5 Flash | 150 | 34 | 22.67% |
| *Overall Dataset* | *450* | *142* | *31.56%* |

## 5. DISCUSSION

This study provides a large-scale empirical examination of vibe-coded tools in a safety-critical, and regulated domain. By generating 450 Python scripts from three frontier LLMs (Claude 3.5 Haiku, GPT-4o-Mini, and Gemini 2.5 Flash) across LLM generated realistic construction personas, executing them in a controlled sandbox, and evaluating them with a validated LLM-as-a-Judge rubric, the findings offer a detailed picture of how contemporary models behave when non-technical users attempt to automate OSHA-compliant safety logic.

A core theoretical contribution of this study is the identification and quantification of the illusion of competence in LLM code generation. Historically, software engineering has treated execution viability, i.e., code that compiles and runs without crashing, as a primary indicator of code quality. However, our findings demonstrate that in domain-specific, cyber-physical applications, high execution viability can actually become a liability.

The models that exhibited the highest syntactical reliability (e.g., GPT-4o-Mini) simultaneously demonstrated the highest propensity for producing structurally brittle, mathematically dangerous outputs. Because the code almost always compiled perfectly, it actively masked dangerous miscalculations in OSHA logic. By formally introducing the Silent Failure Rate (SFR), this study provides a new theoretical lens for LLM evaluation. Syntactic perfection creates a false sense of security, forcing non-technical users to trust outputs they cannot mathematically verify.

Furthermore, this study establishes that in a vibe-coding paradigm, the linguistic register of the user dictates the deterministic safety of the system. The stark contrast in hallucination rates between the Safety Manager and the field-level personas (Foreman and Worker) proves that prompt context is not merely cosmetic. Informality and implied stress inherently degrade the model's adherence to safety boundaries, causing it to invent missing variables rather than halt execution. This contributes to the broader human-computer interaction (HCI) discourse by proving that vibe coding inherits the cognitive load and communication deficits of the user's persona.

The findings of this study have several practical implications for construction safety practice and for the design of AI-enabled tools. First, they suggest that allowing frontline workers or supervisors to independently vibe code safety tools or software using general-purpose LLMs is incompatible with the zero-tolerance posture that safety engineering typically requires. The combination of high hallucination rates under realistic personas, weak architectural robustness, and nearly half of executable scripts failing OSHA logic means that raw LLM outputs should not be treated as deployable safety instruments, even when they appear to function correctly. This risk is severely compounded by the models' frequent failure to provide actionable safety directives. Because over 31% of the generated tools output contextless raw numbers rather than explicit 'SAFE/UNSAFE' warnings, the end-user is forced to become a regulatory interpreter on the fly. This UI/UX failure strips away the intended protective utility of the software, drastically increasing the likelihood that a worker will accept a contextually dangerous output as a verified engineering truth. Organizations that choose to adopt AI for safety applications will need to impose strict boundaries on how LLMs are used, including prohibiting unvetted, one-off scripts from entering production workflows.

Second, the results reinforce the importance of deterministic safety wrappers around probabilistic models. The study's evidence supports a layered architecture in which LLMs are limited to tasks such as parsing natural language, extracting variables from informal descriptions, and mapping user intents to predefined tool functions, while the underlying safety calculations are performed by pre-audited, deterministic logic engines. Under this approach, OSHA formulas and thresholds are encoded once, reviewed and validated by experts, and then invoked by LLM-driven interfaces without allowing the models to improvise the mathematics. Given that even the most syntactically reliable models in this study produced incorrect OSHA logic in over half of their executable scripts, relocating mathematical responsibility to verified code is a necessary control, not an optional enhancement.

Third, the persona results indicate that prompt engineering alone is unlikely to solve the problem at the point of use. While it is possible to train or script workers to provide more structured inputs, the very scenarios where vibe coding is most attractive, time pressure, incomplete information, and informal communication, are also those where hallucinations were most frequent. This suggests that organizational safeguards should focus less on expecting perfect prompts from end-users and more on constraining the system's degrees of freedom, enforcing clear boundaries on what the LLM is allowed to decide, and mandating secondary checks (e.g., comparison against known safe ranges) before any safety recommendation is surfaced.

This study shows that vibe coding in its current, unconstrained form is poorly aligned with the demands of construction safety engineering. The core contribution is not simply that LLMs sometimes make mistakes, but that under realistic user personas, highly capable models frequently produce tools that look syntactically perfect yet embed unsafe assumptions and flawed OSHA logic at scale. For practitioners and policymakers, the implication is clear, LLMs can play a valuable role in assisting with safety documentation, information retrieval, and interface design, but they should not be granted unilateral authority to generate or update the mathematical backbone of safety-critical calculations without supervision. For AI and software engineering researchers, the findings underscore the urgency of developing tools, frameworks, and governance mechanisms that treat silent failures not as an edge case, but as a central design constraint for any AI system operating in high-risk, cyber-physical environments.

## 6. CONCLUSION, LIMITATION, AND FUTURE WORK

This study conducted a large-scale empirical assessment of vibe-coded construction safety tools generated by contemporary frontier LLMs. By simulating realistic jobsite communication styles through persona-driven prompts, executing the generated code in a controlled sandbox, and utilizing a validated LLM-as-a-judge rubric, the research uncovered critical vulnerabilities in zero-shot vibe code generation for high-risk domains such as the construction industry.

The findings converge on three core insights. First, the operational context of the user is a primary determinant of AI safety behavior. When prompted with the informal, high-stress communication typical of frontline construction workers, the models' propensity to hallucinate missing safety-critical variables escalated dramatically compared to formal, structured inquiries. Second, frontier models exhibit a dangerous illusion of architectural competence. While the generated code routinely compiled and executed flawlessly, this syntactic success actively masked brittle underlying architectures that were largely devoid of defensive programming or basic error handling. Third, and most alarmingly, syntactic viability is a remarkably poor proxy for mathematical safety. A substantial portion of the code that executed without crashing silently violated OSHA thresholds or failed to provide actionable, human-readable safety directives, leaving end-users with dangerous miscalculations masquerading as functional software.

These results demonstrate that unconstrained vibe coding introduces an unacceptable risk profile for construction safety engineering. Because the failure modes of generative models are inherently deceptive, masking dangerous logical errors beneath flawlessly executing syntax, treating raw, LLM-generated scripts as deployable tools actively subverts the auditing capabilities of non-technical personnel and violates established principles of occupational hazard control.

To harness the utility of LLMs without compromising jobsite safety, a fundamental architectural paradigm shift is required. A defensible implementation of AI in construction safety must abandon autonomous vibe coding in favor of layered, deterministic architectures. LLMs should be strictly confined to language-facing roles, such as interpreting informal descriptions, extracting variables from user prompts, or navigating dense regulatory text. However, all safety critical calculations must be delegated to pre-audited, deterministic logic engines. By treating the LLM as a natural language interface rather than an autonomous software engineer, the construction industry can leverage the accessibility of generative AI without outsourcing the core protective logic of safety engineering to probabilistic, unverified systems.

While the study offered impactful insights and contributions, several limitations of the present work constrain how its findings should be generalized and point to directions for future research. The evaluation considered three specific models at particular points in time. As LLM architectures and training data evolve, both failure modes and baseline robustness may change over time, even if the qualitative pattern of risk remains similar. The personas were carefully designed but synthetic, and real-world language on construction sites varies across firms, trades, and regions, so actual hallucination and failure rates could differ somewhat from those reported here. All experiments used Python scripts executed in a controlled sandbox, without front-end interfaces, organizational approval workflows, or integrated testing pipelines that might either mitigate or compound risks in deployed systems. The regulatory scope was limited to OSHA 29 CFR 1926, leaving open how models would behave under other standards or jurisdictions. Finally, while the LLM-as-a-Judge rubric achieved almost perfect agreement with human raters during validation, it remains a model-based instrument and may miss nuances of partial correctness or edge-case safety that matter in practice.

Future work should therefore extend this methodology along several axes. One line of research is to repeat and broaden the evaluation across additional models and versions to test the stability of persona effects, architectural fragility, and SFR over time. Another is to move beyond zero-shot prompting and systematically assess whether techniques such as few-shot examples, chain-of-thought prompting, tool calling, or retrieval-augmented generation grounded in OSHA manuals can materially reduce hallucinations and logic errors. Human-in-the-loop experiments with practicing foremen and safety managers could quantify how much expert review and revision is required to bring failure rates within acceptable bounds, and which review workflows are most effective. Finally, applying similar evaluation frameworks to other high-risk domains, such as electrical work, mining, or process industries, and to regulatory environments outside the United States would help distinguish domain-specific issues from more general properties of vibe-coded systems. Such work can inform the design of AI-assisted safety tools that harness LLMs' strengths while embedding them within deterministic, rigorously engineered safety architectures.